\documentclass[draftclsnofoot,12pt, onecolumn]{IEEEtran}
\usepackage{amsmath}
\usepackage{epsfig}
\usepackage{amssymb}
\usepackage{latexsym}
\usepackage{graphicx}
\usepackage{cite}

\title{A Generalized Non-Linear Composite Fading Model}
\author{\IEEEauthorblockN{Paschalis C. Sofotasios}
\IEEEauthorblockA{\\School of Electronic and Electrical Engineering \\
University of Leeds, UK\\
e-mail: p.sofotasios@leeds.ac.uk\\}
\and
\IEEEauthorblockN{Steven Freear\\}
\IEEEauthorblockA{School of Electronic and Electrical Engineering \\ 
University of Leeds, UK\\
e-mail: s.freear@leeds.ac.uk}}
\begin{document}
\maketitle
\begin{abstract} 
This work is devoted to the formulation and derivation of the $\alpha{-}\kappa{-}\mu{/}$gamma distribution which corresponds to a physical fading model. The proposed distribution is composite and is constituted by the $\alpha{-}\kappa{-}\mu$ non-linear generalized multipath model and the gamma shadowing model. It also constitute the basis for deriving the $\alpha{-}\kappa{-}\mu$ \textit{Extreme}${/}$gamma model which accounts for non-linear severe multipath and shadowing effects and also includes the more widely known $\alpha{-}\mu$ and $\kappa{-}\mu$ models which includes as special cases the Rice, Weibull, Nakagami-$m$ and Rayleigh distributions. The derived models provide accurate characterisation of the simultaneous occurrence of multipath fading and shadowing effects. This is achieved thanks to the remarkable flexibility of their named parameters which have been shown to render them capable of providing good fittings to experimental data associated with realistic communication scenarios. This is also evident  by the fact that they include as special cases the widely known composite fading models such as the recently reported $\kappa{-}\mu{/}$gamma model and the novel $\alpha{-}\mu{/}$gamma model. Novel analytic expressions are derived for the corresponding probability density function of these distributions which are expressed in a convenient algebraic form and can be efficiently utilized in the derivation of numerous vital measures in investigations related to the analytic performance evaluation of digital communications over composite multipath${/}$shadowing fading channels.
\end{abstract}

%
%
%
\section{Introduction}
\indent
It is widely known that fading is an effect that significantly degrades communication signals during wireless propagation. A common approach for accounting for this phenomenon has been possible through exploitation of suitable statistical distributions. To this effect, statistical models such as Rayleigh, Nagakami-$m$, Weibull and Nakagami-q (Hoyt) have been shown to be capable of modelling small-scale fading in \textit{non-line-of-sight }(NLOS) communication scenarios whereas Nakagami-$n$ (Rice) distribution has been typically utilized in characterizing multipath fading in \textit{line-of-sight }(LOS) communication scenarios, \cite{B:Nakagami, B:Jakes, B:Alouini} and the references therein. Capitalizing on   these models, M. D. Yacoub proposed three generalised fading models, namely, the $\alpha{-}\mu$, the $\kappa{-}\mu$, the $\eta{-}\mu$ models and subsequently the $\lambda{-}\mu$ and the $\kappa{-}\mu$ \textit{Extreme} models, \cite{C:Yacoub_1, C:Yacoub_2, C:Yacoub_3, J:Yacoub_1, J:Yacoub_2, C:Yacoub_4, C:Rabelo}. These models along with the recently proposed $\alpha{-}\kappa{-}\mu$ \textit{Extreme} fading model in \cite{C:Sofotasios_1} are useful thanks to the remarkable flexibility offered by their named parameters which render them capable of providing adequate fittings to results obtained by field measurements. Their usefulness is also evident by the fact that they include as special cases all the above small-scale fading distributions. \\
\indent
It is recalled here that a fundamental principle of wireless radio propagation is that multipath and shadowing effects occur simultaneously. As a result, in spite of the undoubted usefulness of the aforementioned fading models, they all ultimately fail to account concurrently for both shadowing and multipath fading. In other words, the characterisation offered by the aforementioned fading models is limited to modelling either the one or the other effect. Based on this crucial limitation, the need for composite statistical models that can provide adequate characterization of the fading effect as a whole became necessary \cite{B:Nakagami, B:Jakes, B:Alouini}. \\
\indent
Motivated by this, the authors in \cite{J:Kaveh} proposed the Rayleigh${/}$gamma fading model, which is also known as $K$-distibution, $(K)$.  In the same context, Shankar in \cite{J:Shankar} exploited the flexibility of Nakagami-$m$ distribution, which includes Rayleigh distribution as a special case and introduced the Nakagami-$m{/}$gamma composite distribution - or generalised $K$-distribution, $(K_{G})$. Likewise, the Weibull${/}$gamma composite distribution was reported in \cite{J:Bithas} while the $\kappa{-}\mu{/}$gamma, the $\kappa{-}\mu$ \textit{Extreme}${/}$gamma, the $\eta{-}\mu{/}$gamma and the $\lambda{-}\mu {/}$gamma distributions were proposed  in \cite{C:Sofotasios_2, C:Sofotasios_3, C:Sofotasios_4, C:Sofotasios_5, C:Sofotasios_6, B:Sofotasios}. \\
\indent
The aim of this work is the derivation of novel analytic results for the $\alpha{-}\kappa{-}\mu{/}$gamma composite fading model. Subsequently, by using this model as a basis, the $\alpha{-}\mu{/}$gamma and the $\alpha{-}\kappa{-}\mu$ \textit{Extreme}${/}$gamma composite fading distributions are additionally derived. Importantly, unlike all existing composite fading models, this models are capable of accounting for the non-linearity of the wireless fading channel which is characterized by the parameter $\alpha$. After formulating these models, novel analytic expressions are derived for their corresponding probability density function (pdf).  The validity of the offered expressions is justified through comparisons with numerical results while their behaviour is assessed under different parametric scenarios. Importantly, owing to the relatively convenient algebraic representation of the offered expressions, they can be considered useful mathematical tools that can be efficiently utilized in studies related to the analytical performance evaluation of digital communications over non-linear multipath${/}$shadowing and severe composite fading models. Indicatively, they can be exclusively employed in the derivation of algebraically complex analytic expressions for various performance measures such as error probability, channel capacity and higher order statistics, to name a few \cite{New_0, New_1, New_2, New_3, New_4, New_5}.\\
\indent
The remainder of this paper is organised as follows: Section II revisits the foundations of $\alpha{-}\kappa{-}\mu$ and the gamma distributions. Subsequently, Sections III, IV and V are devoted to the formulation and analytical derivation of the $\alpha{-}\kappa{-}\mu{/}$gamma, the $\alpha{-}\mu{/}$gamma and the $\alpha{-}\kappa{-}\mu$ \textit{Extreme}${/}$gamma fading models, respectively. Finally, discussions on the potential applicability of the derived expressions in wireless communications along with closing remarks are given in Section VI.  
%
%
\section{The $\alpha{-}\kappa{-}\mu$ and Gamma Fading Distributions}
\subsection{The $\alpha{-}\kappa{-}\mu$ Fading Model}

The $\alpha{-}\kappa{-}\mu$ distribution is a recently proposed fading model that accounts for the characterization of small-scale variations of mobile radio signals in LOS communication scenarios. It is written in terms of three physical parameters, namely $\alpha$, $\kappa$ and $\mu$; the first one denotes the non-linearity parameter, $\mu$ is related to the multipath clustering while $\kappa$ represents the ratio between the total power of the dominant components and the total power of the scattered waves. For a fading signal with envelope $R$ and $\hat{r} = \,^{a}\sqrt{E(R^{a})}$, the $\alpha{-}\kappa{-}\mu$ envelope pdf is given by \cite{C:Yacoub_5}, 

\begin{equation} \label{eq:one} 
p_{R}(r) = \frac{\alpha \mu I_{\mu - 1}\left(2\mu \sqrt{\kappa(1 + \kappa)}\frac{r^{\alpha/2}}{\hat{r}^{\alpha/2}} \right)}{ \hat{r}(1+\kappa)^{-\frac{1+\mu}{2}}\kappa^{\frac{\mu -1}{2}}e^{\mu\left(\kappa + \frac{r^{a}}{\hat{r}^{\alpha}} + \kappa \frac{r^{\alpha}}{\hat{r}^{\alpha}} \right)}} \left(\frac{r}{\hat{r}} \right)^{\frac{\alpha(1+\mu)}{2} - 1}
\end{equation}
where $E(.)$ denotes expectation, $\hat{r}$ is the root-mean-square $(rms)$ value of $R$ and $I_{\nu}(x)$ represents the modified Bessel function of the first kind with argument $x$ and order $\nu$ \cite{B:Tables}. 
For the case of normalised envelope $P = R/\hat{r}$, equation \eqref{eq:one} can be equivalently expressed as

\begin{equation} \label{eq:two} 
p_{P}(\rho) = \frac{\alpha \mu  \rho^{\frac{\alpha(1+\mu)}{2} - 1}I_{\mu - 1}\left(2\mu \sqrt{\kappa(1 + \kappa)} \rho^{\alpha/2} \right)}{\kappa^{\frac{\mu -1}{2}}(1+\kappa)^{-\frac{1+\mu}{2}}e^{\mu\left(\kappa + \rho^{\alpha}+ \kappa \rho^{\alpha} \right)}}
\end{equation}
while the corresponding cdf is expressed as,

\begin{align} \label{eq:three} 
F_{P}(\rho) =& \sum_{i=0}^{\infty} \frac{\kappa^{i}\mu^{i}\Gamma\left(i + \mu, \mu (1+\kappa)\rho^{\alpha} \right)}{i!e^{\mu \kappa}\Gamma(i + \mu)} \\ =& 1 - Q_{\mu}\left(\sqrt{2\mu \kappa}, \rho^{\frac{\alpha}{2}}\sqrt{2\mu(1+\kappa)} \right)
\end{align}
where $\Gamma(a,x)$ and $Q_{m}(a,b)$ are the upper incomplete Gamma function and the generalized Marcum $Q$-function which are defined as \cite{B:Tables,B:Abramowitz}. Importantly, the $\alpha{-}\kappa{-}\mu$ fading model includes as specific cases the more widely known $\alpha{-}\mu$ and $\kappa{-}\mu$ generalized models \cite{C:Yacoub_5}. In more details, for the specific case that $\alpha = 2$, the $\alpha{-}\kappa{-}\mu$ distribution reduces to the more widely known $\kappa{-}\mu$ fading model. Likewise, for the specific case $\alpha=2$ and $\kappa=0$, the Nakagami-$m$ distribution is deduced while for $m=1$ one obtains the Rayleigh distribution. In the same context, the Rice distribution can be obtained by setting $\kappa=k$ (where k denotes the Rice parameter) as well as for the specific case $\alpha=2$ and $\mu=1$. Finally, the Weibull distribution can be obtained by setting $\kappa = 0$ and $\mu = 1$. \\
\indent
By setting $p_{W}(w) = p_{P}(\sqrt{w})/2\sqrt{w}$ according to \cite{J:Yacoub_2}, the corresponding power (pdf) is deduced in a straightforward manner, namely, 

\begin{equation} \label{eq:four} 
p_{P}(w) = \frac{\alpha \mu  w^{\frac{\alpha(1+\mu)}{4} - 1}I_{\mu - 1}\left(2\mu \sqrt{\kappa(1 + \kappa) w^{\alpha/2}} \right)}{2\kappa^{\frac{\mu -1}{2}}(1+\kappa)^{-\frac{1+\mu}{2}}e^{\mu\left(\kappa + w^{\alpha/2}+ \kappa w^{\alpha/2} \right)}}
\end{equation}
In the same context, the corresponding statistical moments are given explicitly by,

\begin{equation} \label{eq:five} 
E\left[P^{l} \right] = \frac{\Gamma\left(\frac{l}{\alpha} + \mu \right)\,_{1}F_{1}\left(\frac{l}{\alpha}, \mu, \kappa \mu \right)}{e^{\mu \kappa}(1+\kappa)^{l/ \alpha}\mu^{l/ \alpha}}\Gamma(\mu)
\end{equation}
where the notation  $\,_{1}F_{1}(a;b;x)$ denotes the Kummer confluent hypergeometric function, \cite{B:Tables}. \\
\indent
Finally, it is recalled that the Nakagami parameter $m$ is regarded as the inverse of the variance of the normalized power of the fading signal, i.e. $m=Var^{-1}\left(P^{2} \right)$, \cite{J:Yacoub_2}. This term is linked with the $\kappa$ and $\mu$ parameters through the following relationship, 

\begin{equation}\label{eq:6} 
m = \frac{\mu (1+\kappa)^{2}}{1 + 2\kappa}
\end{equation}

As already mentioned, the $\alpha{-}\kappa{-}\mu$ fading model includes as a special case the $\alpha{-}\kappa{-}\mu$ \textit{Extreme} distribution which subsequently includes the $\kappa{-}\mu$ \textit{Extreme} model for $\alpha=2$. It is recalled that the validity of this model is based on that  the occurrence of  large number of paths is not typically the case in wireless radio propagation over enclosed environments such as aeroplanes, trains and buses \cite{J:Frolik, J:Durgin}. The reason underlying this principle is that such environments are known to be characterised by only a few number of paths. As a consequence, severe fading conditions, even worse than Rayleigh, are ultimately constituted \cite{C:Rabelo, J:Frolik, J:Durgin}. The envelope pdf of the $\kappa{-}\mu$ fading model was derived in \cite{C:Rabelo} and is given by,

\begin{equation}  \label{eq:7} 
p_{P}(\rho) = 4m e^{-2m\left(1 + \rho^{2} \right)}I_{1}\left(4m\rho \right) + e^{-2m} \delta(\rho)  
\end{equation}
while  the $\alpha{-}\kappa{-}\mu$ model which additionally includes the non-linearity of the wireless channel was recently proposed in \cite{C:Sofotasios_5} and is defined as follows,

\begin{equation}  \label{eq:8} 
p_{P}(\rho) = \frac{2\alpha m I_{1}\left(4m\rho^{\alpha/2} \right)}{\rho^{1-\frac{\alpha}{2}}e^{2m\left(1 + \rho^{\alpha} \right)}} + e^{-2m} \delta(\rho) 
\end{equation}
For $\alpha = 2$ equation \eqref{eq:8} reduces to \eqref{eq:7}. 
\subsection{The Gamma Fading Model}
\indent
It is recalled that log-normal distribution has been regarded the optimum statistical model for characterising the shadowing effect, \cite{B:Nakagami, B:Jakes, B:Alouini}. Nevertheless, in spite of its usefulness, it has been largely shown that when it is involved in combination with other elementary and/or special function, its algebraic representation becomes intractable. This is particularly the case in studies related to the analytical derivation of critical performance measures in the field of digital communications over fading channels. Motivated by this, the authors in \cite{J:Kaveh} proposed the gamma distribution as an accurate substitute to log-normal distribution. Mathematically, the envelope pdf of gamma distribution is given by \cite[eq. (4)]{J:Kaveh}, namely,

\begin{equation} \label{eq:9} 
p_{_{Y}}(y) = \frac{y^{b-1}e^{-\frac{y}{\Omega}}}{\Gamma(b) \Omega^{b}}, \qquad \, \, y\geq 0 
\end{equation}
where the term $b > 0$ is its shaping parameter and $\Omega = E(Y^{2})$. The gamma fading model has been shown to provide adequate fitting to experimental data that correspond to realistic fading conditions. In addition, it is evident that its algebraic representation is quite tractable and therefore, easy to handle both analytically and numerically. As a result, it has been undoubtedly useful in characterising shadowing and for this reason it has been exploited in the formulation of the $K$ and $K_{G}$ composite multipath/shadowing models, \cite{J:Kaveh, J:Shankar}. 
%
\section{The $\alpha{-}\kappa{-}\mu{/}$gamma Fading Distribution}
\subsection{Model Formulation}
\indent
According to the basic principles of statistics, the envelope pdf of a composite statistical distribution is deduced by the superposition of two or more statistical distributions. In the present case, this is realized by superimposing one multipath and one shadowing distribution. Mathematically,  this principle is expressed as, 

\begin{equation} \label{eq:10} 
p_{_{X}}(x) = \int_{0}^{\infty} p_{_{X\mid Y}}(x\mid y)p_{_{Y}}(y)dy
\end{equation}
where $p_{_{X\mid Y}}(x\mid y)$ denotes the corresponding multipath distribution with mode $y$. Based on this principle, the corresponding $\kappa-\mu$/gamma composite fading distribution is formulated by firstly setting in \eqref{eq:one}, $r = x$ and $\hat{r}=y$  and then substituting  in \eqref{eq:10} along with equation \eqref{eq:9}. To this end, it follows that,

\begin{equation} \label{eq:11} 
p_{_{X}}(x) = \mathcal{A} \int_{0}^{\infty} \frac{I_{\mu-1}\left(2\mu \sqrt{\kappa (1 + \kappa)} \frac{x^{\alpha /2}}{y^{\alpha /2}} \right)}{y^{1  + \frac{\alpha (1 + \mu)}{2}- b} e^{\mu (1 + \kappa) \frac{x^{a}}{y^{a}}}e^{\frac{y}{\Omega}}} dy
\end{equation}
where

\begin{equation}
\mathcal{A} =\frac{\alpha \mu x^{\frac{\alpha(1 + \mu)}{2}-1} (1 + \kappa)^{\frac{1 + \mu}{2}} }{e^{\mu \kappa}\kappa^{\frac{\mu -1}{2}}\Gamma(b) \Omega^{b}}
\end{equation}

By setting in \eqref{eq:11}, $u = y^{\alpha}$ and thus, $du/dy = \alpha y^{\alpha -1}$ and $y = u^{1/\alpha}$, the above expression can be equivalently re-written as follows:
 
\begin{equation} \label{eq:12} 
p_{_{X}}(x) = \frac{\mathcal{A} }{\alpha} \int_{0}^{\infty} \frac{I_{\mu-1}\left(2\mu \sqrt{\kappa (1 + \kappa)} \frac{x^{\alpha /2}}{\sqrt{u}} \right)}{u^{1 - \frac{b}{\alpha} + \frac{\mu  + 1}{2} } e^{\mu (1 + \kappa) \frac{x^{\alpha}}{u}}e^{\frac{u^{1/\alpha}}{\Omega}}} du
\end{equation}

\subsection{A Novel Expression for the Envelope pdf}

It is evident from \eqref{eq:12} that the derivation of an analytic expression for the envelope pdf of the $\alpha{-}\kappa{-}\mu{/}$gamma distribution is subject to evaluation of the involved integral. To this end, it is recalled the the $I_{\nu}(x)$ function can be equivalently represented in terms of the infinite series in \cite[eq. (8.445)]{B:Tables}, and the polynomial approximation in \cite[eq. (19)]{J:Gross}, namely,

\begin{align} \label{eq:13} 
I_{\nu} (x) =& \sum_{l = 0}^{\infty} \frac{1}{\Gamma(l + 1) \Gamma(\nu + l + 1)} \left(\frac{x}{2} \right)^{\nu + 2l} \\
 \simeq &\sum_{l = 0}^{n} \frac{\Gamma(n + l)}{\Gamma(l + 1) \Gamma(n - l + 1)}\frac{n^{1 - 2l}}{\Gamma(\nu + l + 1)} \left(\frac{x}{2} \right)^{\nu + 2l}
\end{align}
Therefore, by making the necessary change of variables\footnote{The polynomial approximation is used in the present analysis.}, it immediately follows that\footnote{As $n \rightarrow \infty$, the polynomial approximation reduces to the infinite series.}

\begin{equation} \label{eq:14} 
I_{\mu - 1}\left(2\mu \sqrt{\kappa (1 + \kappa)}\frac{x^{\alpha / 2}}{\sqrt{u}} \right) \simeq  \sum_{l = 0}^{n}\frac{\Gamma(n + l)n^{1-2l}x^{\frac{\alpha}{2} + \mu +2l -1}\mu^{\mu + 2l - 1}\kappa^{\frac{\mu -1}{2} + l}(1 + \kappa)^{\frac{\mu -1}{2}+l}}{\Gamma(l+1)\Gamma(n -l +1)\Gamma(\mu +l)u^{\frac{\mu -1}{2}+l}}
\end{equation}
Subsequently, by recalling that $\Gamma(x)\triangleq (x-1)!$ and substituting in \eqref{eq:14} in \eqref{eq:12} yields,

\begin{equation}  \label{eq:15} 
p_{_{X}}(x) =  \sum_{l = 0}^{n} \frac{\Gamma(n + l)x^{\alpha (\mu + l) - 1 }\mu^{\mu + 2l} \kappa^{l}(1 + \kappa)^{\mu + l}}{l!\Gamma(n - l + 1) \Gamma(\mu + l) \Gamma(b) \Omega^{b}n^{2l - 1}e^{\mu \kappa}} \underbrace{\int_{0}^{\infty} u^{\frac{b}{\alpha} - \mu - l - 1}e^{-\mu (1 + \kappa) \frac{x^{a}}{u}} e^{-\frac{u^{1/\alpha}}{\Omega}} du}_{\mathcal{I}_{1}}
\end{equation}
Importantly, the $\mathcal{I}_{1}$ integral in \eqref{eq:15} can be solved in closed-form. To this end, by expressing the corresponding exponentials in terms of the Meijer G-function according to \cite[eq. (8.4.3.1)]{B:Brychkov} and  \cite[eq. (8.4.3.2)]{B:Brychkov}, yields,

\begin{equation} \label{eq:16} 
\mathcal{I}_{1} = \int_{0}^{\infty} \frac{         G_{0,1}^{1,0}\left( \frac{u^{1/\alpha}}{\Omega}\Big| _{0}^{.} \right)               G_{1,0}^{0,1}\left(\frac{u}{\mu (1 + \kappa) x^{\alpha}} \Big| _{.}^{1}  \right)}{u^{1 + \mu + l  - \frac{b}{\alpha} }      } du
\end{equation}
By setting in \eqref{eq:16}, $z = u^{1/\alpha}$ and thus, $u = z^{\alpha}$ and $dz/du = u^{\frac{1}{\alpha}-1}/\alpha$, one obtains, 

\begin{equation}  \label{eq:17} 
\mathcal{I}_{1} = \alpha \int_{0}^{\infty} \frac{         G_{0,1}^{1,0}\left( \frac{z}{\Omega}\Big| _{0}^{.} \right)               G_{1,0}^{0,1}\left(\frac{z^{\alpha}}{\mu (1 + \kappa) x^{\alpha}} \Big| _{.}^{1}  \right)}{z^{\alpha (\mu + l) - b + 1 }      } du
\end{equation}
In its current form, the above expression can be expressed in closed-form according to \cite[eq. (2.24.1.1)]{B:Brychkov} yielding, 

\begin{equation} \label{eq:18} 
\mathcal{I}_{1} = \frac{\alpha^{b - \alpha(\mu + l) +1}}{(2\pi)^{\frac{\alpha - 1}{2}}\Omega^{\alpha(\mu + l) - b}}  G_{1+\alpha, 0}^{0,1+\alpha}\left( \frac{\Omega^{\alpha} \alpha^{\alpha}}{\mu (1 + \kappa)x^{\alpha}}\left| _{\,\,\,\,\, \,\, \frac{1 - b + \alpha(\mu + l)}{l}}^{1, 1, \frac{1 - b + \alpha(\mu + l)}{l}}\right. \right)   
\end{equation}
Therefore, by substituting \eqref{eq:18} in \eqref{eq:15}, an analytic expression for the envelope pdf of the $\alpha{-}\kappa{-}\mu{/}$gamma composite fading model is deduced, namely, 

\begin{equation}  \label{eq:19} 
p_{_{X}}(x) = \sum_{l = 0}^{n} \frac{\Gamma(n + l)x^{\alpha (\mu + l) - 1 }\mu^{\mu + 2l} \kappa^{l}(1 + \kappa)^{\mu + l}}{l!\Gamma(n - l + 1) \Gamma(\mu + l) \Gamma(b) n^{2l - 1}e^{\mu \kappa}}  \frac{\alpha^{b - \alpha(\mu + l) +1}}{(2\pi)^{\frac{\alpha - 1}{2}}\Omega^{\alpha(\mu + l)}}  G_{1+\alpha,0}^{0,1+\alpha}\left( \frac{\Omega^{\alpha} \alpha^{\alpha}}{\mu (1 + \kappa)x^{\alpha}}\left| _{\,\,\,\,\, \,\, \frac{1 - b + \alpha(\mu + l)}{l}}^{1, 1, \frac{1 - b + \alpha(\mu + l)}{l}}\right. \right)  
\end{equation}
which to the best of the authors' knowledge, it has not been previously reported in the open technical literature. 
\subsection{An analytic expression for the pdf of $\kappa{-}\mu{/}$gamma model}
As already mentioned, analytic expressions for the pdf of the $\kappa{-}\mu{/}$gamma distribution was reported recently in \cite{C:Sofotasios_2, C:Sofotasios_5}. By recalling   that the $\kappa{-}\mu$ distribution is a special case of the $\alpha{-}\kappa{-}\mu$ distribution, an additional analytic expression for the pdf of $\kappa{-}\mu{/}$gamma distribution can be obtained by setting $\alpha = 2$ in \eqref{eq:19}. Therefore, one obtains straightforwardly, 

 \begin{equation}  \label{eq:20} 
p_{_{X}}(x) = \sum_{l = 0}^{n} \frac{\Gamma(n + l)x^{2 (\mu + l) - 1 }\mu^{\mu + 2l} \kappa^{l}(1 + \kappa)^{\mu + l}}{l!\Gamma(n - l + 1) \Gamma(\mu + l) \Gamma(b) n^{2l - 1}e^{\mu \kappa}}  \frac{2^{b - 2(\mu + l) +\frac{1}{2}}}{\sqrt{\pi}\Omega^{2(\mu + l)}}  G_{3,0}^{0,3}\left( \frac{4\Omega^{2} }{\mu (1 + \kappa)x^{2}}\left| _{\,\,\,\,\, \,\, \frac{1 - b + 2(\mu + l)}{l}}^{1, 1, \frac{1 - b + 2(\mu + l)}{l}}\right. \right)  
\end{equation}
The above expression could be useful in cases that the analytic expressions in \cite{C:Sofotasios_2, C:Sofotasios_5} are analytically intractable since although the Meijer G-functions are considered rather laborious, they commonly allow the derivation of corresponding solutions in closed-form. 
%
%
\section{The $\alpha{-}\mu{/}$gamma Composite Fading Model. }
\subsection{Model Formulation}
Unlike reported results in generalised linear and non-linear multipath${/}$shadowing fading models, no results have been reported on composite models that are based on the $\alpha{-}\mu$ fading distribution. Motivated by this, this sub-section is devoted to the derivation of a closed-form expression for the corresponding envelope pdf. It is firstly recalled that the $\alpha{-}\mu$ distribution is a generalised small-scale fading model which also accounts for the non-linearity of the wireless channel. It was reported in  \cite{C:Yacoub_1} and it includes as special cases the well known Weibull, Nakagami-$m$ and Rayleigh distributions. 

Its envelope pdf is expressed as

\begin{equation} \label{eq:21} 
p_{_{R}}(r) = \frac{\alpha \mu^{\mu}r^{\alpha \mu - 1}}{\hat{r}^{\alpha \mu} \Gamma(\mu)}e^{-\mu \frac{r^{\alpha}}{\hat{r}^{\alpha}}}
\end{equation}
The $\alpha{-}\mu{/}$gamma distribution is formulated by taking the conditional probability, $p_{_{R}}(r\mid y)$ and averaging over the statistics of the gamma distribution, namely, 

\begin{align} \label{eq:22} 
p_{_{R}}(r) \triangleq &  \int_{0}^{\infty} p_{_{R}}(r\mid y) p_{y}(y)dy \\
=& \frac{\alpha \mu^{\mu} r^{\alpha \mu - 1}}{\Gamma(\mu)\Gamma(b)\Omega^{b}} \int_{0}^{\infty} y^{b - \alpha \mu -1}e^{-\mu \frac{r^{\alpha}}{y^{\alpha}}} e^{-\frac{y}{\Omega}} dy
\end{align}\label{eq:23} 

\subsection{An exact closed-form expression for the envelope pdf}

It is evident that an analytic expression for the envelope pdf of the $\alpha{-}\mu{/}$gamma fading model is subject to evaluation of the integral in \eqref{eq:22}. To this end, by utilizing again \cite[eq. (8.4.3.1)]{B:Brychkov} and  \cite[eq. (8.4.3.2)]{B:Brychkov}, it follows that,

\begin{equation}
p_{_{R}}(r) = \alpha \mu^{\mu} \int_{0}^{\infty}\frac{ G_{0,1}^{1,0}\left( \frac{y}{\Omega}\left| _{0}^{.} \right. \right)               G_{1,0}^{0,1}\left(\frac{y^{\alpha}}{\mu r^{\alpha}} \left| _{.}^{1} \right.  \right)}{ y^{1 + \alpha \mu -b}  r^{1 - \alpha \mu }\Omega^{b} \Gamma(\mu)\Gamma(b)} dy
\end{equation}
Notably, the above integral has the same algebraic representation as \eqref{eq:17}. Therefore, by utilizing \cite[eq. (2.24.1.1)]{B:Brychkov}, one obtains a closed-form expression for the envelope pdf of the $\alpha{-}\mu{/}$gamma distribution, namely, 

 \begin{equation}  \label{eq:24} 
p_{_{R}}(r) =\frac{\mu ^{\mu}G_{1+\alpha,0}^{0,1+\alpha}\left( \frac{\Omega^{\alpha} \alpha^{\alpha} }{\mu (1 + \kappa)r^{\alpha}}\left| _{\,\,\,\,\, \,\, \frac{1 - b + 2(\mu + \alpha )}{\alpha }}^{1, 1, \frac{1 - b + 2(\mu + \alpha )}{\alpha }}\right. \right)  }{\Gamma(\mu)\Gamma(b)(2\pi)^{\frac{\alpha - 1}{2}}\Omega^{\alpha \mu}\alpha^{\alpha \mu -b -1}} 
\end{equation}
To the best of the authors' knowledge equation \eqref{eq:24} is novel.  
%
%
\section{The $\alpha{-}\kappa{-}\mu$ \textit{Extreme}${/}$gamma distribution.}
\subsection{Model Formulation}
\indent
It is recalled that the $\alpha{-}\kappa{-}\mu$ \textit{Extreme} fading model emerges from the $\alpha{-}\kappa{-}\mu$ distribution and accounts for severe fading conditions that typically occur in enclosed environments. It constitutes a generalization of the $\kappa{-}\mu$ \textit{extreme} distribution since it additionally considers the non-linearity effects of the wireless medium. The basic principle underlying this model is that the CLT is not valid since the number of multipaths is rather low. 

However, likewise the $\alpha{-}\kappa-\mu$ distribution, the $\alpha{-}\kappa{-}\mu$ \textit{Extreme} fading model does not account for the simultaneous occurrence of shadowing. Therefore, by following the same procedure as above, the $\alpha{-}\kappa{-}\mu{/}$\textit{Extreme} gamma distribution is formulated. Subsequently, an analytic expression is derived for the corresponding envelope pdf. To this end, the conditional probability of the $\alpha{-}\kappa{-}\mu$ \textit{Extreme} distribution with mode $y$ is straightforwardly expressed as

\begin{equation} \label{eq:25} 
p_{_{R}}(r\mid y) = 2\alpha m \frac{r^{\frac{\alpha}{2} - 1}}{y^{\frac{\alpha}{2}}e^{2m}} e^{-2m\frac{r^{\alpha}}{y^{\alpha}}}I_{1}\left(4m \frac{r^{\alpha /2}}{y^{\alpha /2}} \right)
\end{equation}
Averaging the above expression over the gamma shadowing statistics yields,

\begin{equation} \label{eq:26} 
p_{_{R}}(r) = \frac{2\alpha m r^{\frac{\alpha}{2} - 1}}{\Gamma(b)\Omega^{b}e^{2m}} \int_{0}^{\infty}\frac{ e^{-2m\frac{r^{\alpha}}{y^{\alpha}}}  }{ y^{1 + \frac{\alpha}{2} - b} e^{\frac{y}{\Omega}}}I_{1}\left(4m \frac{r^{\alpha /2}}{y ^{\alpha/2}} \right) dy
\end{equation}
By expressing the $I_{1}(.)$ function in terms of \eqref{eq:13} and the exponential functions according to \cite[eq. (8.4.3.1)]{B:Brychkov} and  \cite[eq. (8.4.3.2)]{B:Brychkov}, equation \eqref{eq:26} can be equivalent re-written as

\begin{equation} \label{eq:27} 
p_{_{R}}(r) \simeq   \sum_{l=0}^{n} \frac{4\alpha m^{2(1+2l)}r^{\alpha(1+l) - 1}\Gamma(n+l)n^{1 - 2l}}{l!\Gamma(b)\Omega^{b}e^{2m}\Gamma(n-l+1)\Gamma(l+2)} \int_{0}^{\infty} y^{b- a(1 + l) - 1} G_{0,1}^{1,0}\left( \frac{y}{\Omega}\left| _{0}^{.} \right. \right) G_{1,0}^{0,1}\left(\frac{y^{\alpha}}{2m r^{\alpha}} \left| _{.}^{1} \right.  \right) dy
\end{equation}
Finally, with the aid of \cite[eq. (2.24.1.1)]{B:Brychkov}, one obtains the following analytic expression for the envelope pdf of the $\alpha{-}\kappa{-}\mu$ \textit{Extreme}${/}$gamma fading model, 

\begin{equation} \label{eq:28} 
p_{_{R}}(r) \simeq  \sum_{l=0}^{n} \frac{2^{\frac{5 - \alpha}{2}} \alpha^{b - \alpha(l+1)+1} m^{2(1+2l)}r^{\alpha(1+l) - 1}\Gamma(n+l)}{l!\Gamma(b)\Omega^{b}e^{2m}\Gamma(n-l+1)\Gamma(l+2)n^{2l-1}}  G_{1+\alpha,0}^{0,1+\alpha}\left( \frac{\Omega^{\alpha} \alpha^{\alpha} }{\mu (1 + \kappa)r^{\alpha}}\left| _{\,\,\,\,\, \,\, \frac{1 - b + 2(\mu + \alpha )}{\alpha }}^{1, 1, \frac{1 - b + 2(\mu + \alpha )}{\alpha }}\right. \right) 
\end{equation}

\subsection{The pdf of the $\kappa{-}\mu$ \textit{Extreme}${/}$gamma fading model.}
The $\kappa{-}\mu$ \textit{Extreme} ${/}$gamma composite fading model was recently reported in \cite{C:Sofotasios_4, C:Sofotasios_5}. An alternative analytic expression for the corresponding envelope pdf can be straightforwardly deduced by the $\alpha{-}\kappa{-}\mu$ \textit{Extreme}${/}$gamma distribution. This is realised for the specific case that $\alpha=2$. To this effect, it immediately follows that, 
 
\begin{equation} \label{eq:29} 
p_{_{R}}(r) \simeq  \sum_{l=0}^{n} \frac{2^{b - 2l + \frac{1}{2}}  m^{2(1+2l)}r^{2(1+l) - 1}\Gamma(n+l)}{l!\Gamma(b)\Omega^{b}e^{2m}\Gamma(n-l+1)\Gamma(l+2)n^{2l-1}}   G_{3,0}^{0,3}\left( \frac{4 \Omega^{2}  }{\mu (1 + \kappa)r^{2}}\left| _{\,\,\,\,\, \,\, \frac{1 - b}{2} + \mu +2}^{1, 1, \ \frac{1 - b}{2} + \mu +2}\right. \right) 
\end{equation}
%
%
\begin{figure}[h]
\centerline{\psfig{figure=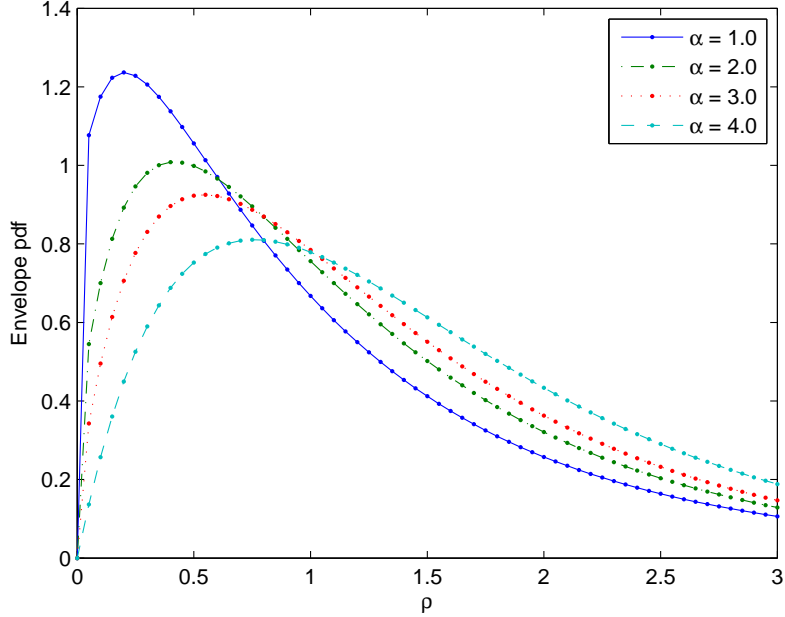,width=12cm, height=9.0cm}}
\caption{Envelope pdf of the $\alpha{-}\kappa{-}\mu{/}$gamma distribution for $b=1.1$, $\Omega=0.9$, $\mu = 2.1$ and different values of $\alpha$. }
\end{figure}
\section{Numerical Results and Discussions}

This Section is devoted to the demonstration of the general behaviour of the derived analytic expressions. Figure $1$ illustrates the envelope pdf of $\alpha{-}\kappa{-}\mu{/}$gamma distribution with respect to $x$ for the case that $b=1.1$, $\Omega=0.9$, $\mu = 2.1$ and different values of $\alpha$. One can clearly observe the flexibility offered by the non-linearity parameter. Likewise, Figure $2$ considers $b=1.8$, $\Omega=0.7$, $\kappa = 4.0$, $\alpha = 2$ and different values of $\mu$. One again can observe the sensitivity of the $\mu$ parameter particularly in the small value regime. 

In the same context, Figure $3$ demonstrates the envelope pdf of the $\alpha{-}\mu{/}$gamma distribution for the general case that $b=1.1$, $\Omega=0.9$, $\mu = 2.1$ and different values of $\alpha$. The characterization flexibility thanks to $\alpha$ is evident and this model is considered ideal for small-scale fading in NLOS communication scenarios. Finally, the behaviour of the envelope pdf of the $\alpha{-}\kappa{-}\mu$ \textit{extreme}${/}$gamma fading model is illustrated in Figure $4$ for $\Omega = 0.8$, $b=1.2$, $m = 1.1$ and different values of $\alpha$. 
\begin{figure}[h]
\centerline{\psfig{figure=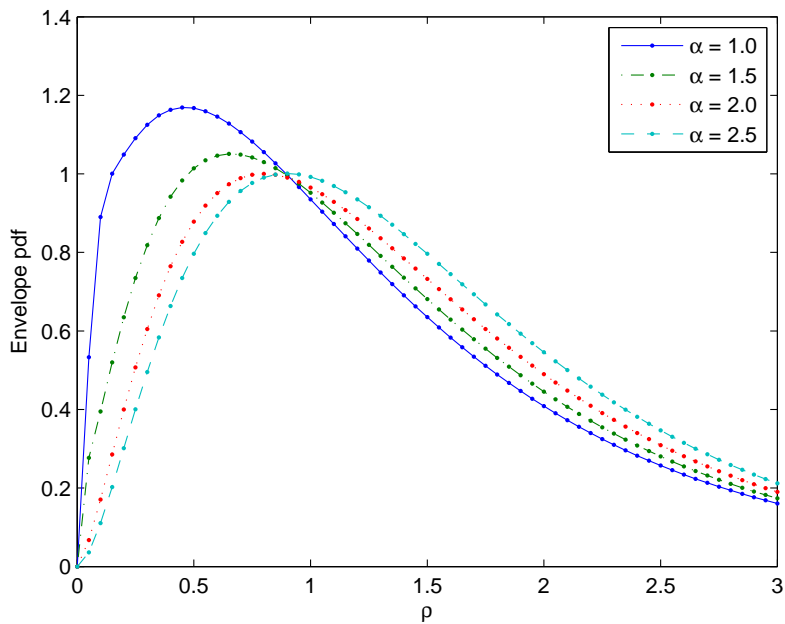,width=12cm, height=9.0cm}}
\caption{Envelope pdf of the $\kappa{-}\mu{/}$gamma distribution for $\alpha = 2$, $b=1.8$, $\Omega=0.7$, $\kappa = 4.0$ and different values of $\mu$}
\end{figure}
\begin{figure}[h]
\centerline{\psfig{figure=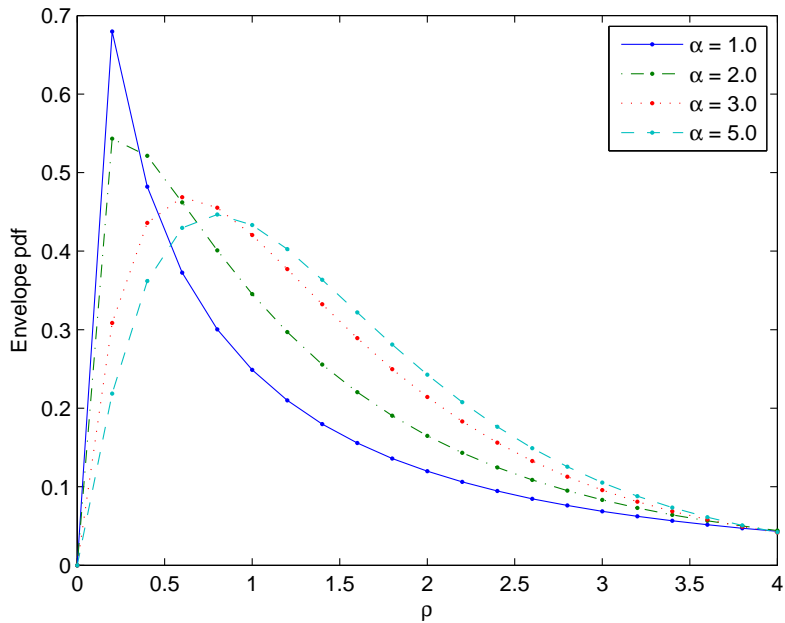,width=12cm, height=9.0cm}}
\caption{Envelope pdf of the $\alpha{-}\mu$ fading model for  $b=1.1$, $\Omega=0.9$, $\mu = 2.1$ and different values of $\alpha$}
\end{figure}
\subsection{Usefulness and Applicability in Wireless Communications}
It is widely known that the algebraic representation of crucial performance measures is rather critical in studies related to analytical performance evaluation of digital communications. This is obvious by the fact that when the algebraic form of a corresponding measure is convenient, it ultimately becomes more possible that the derived relationships can be expressed in closed-form. Based on this, the fact that the form of the offered analytic expressions is suitable due to the existence of various identities involving Meijer G-functions, constitutes the proposed fading models convenient to handle both analytically and numerically. Therefore, the derived expressions can be efficiently applied in various analytic studies relating to the performance evaluation of digital communications over composite multipath/shadowing fading channels including non-linearities as well as severe fading conditions. Indicatively, they can be meaningfully utilized in deriving explicit expressions for important performance measures such as, error probability, probability of outage, ergodic capacity, channel capacity under different transmission policies and higher order statistics. It is recalled here that expressions corresponding to the aforementioned measures can obviously be derived in both classical and emerging technologies such as single channel and multichannel communications, cognitive radio and cooperative systems and free-space optical communications, to name a few.
\begin{figure}[h]
\centerline{\psfig{figure=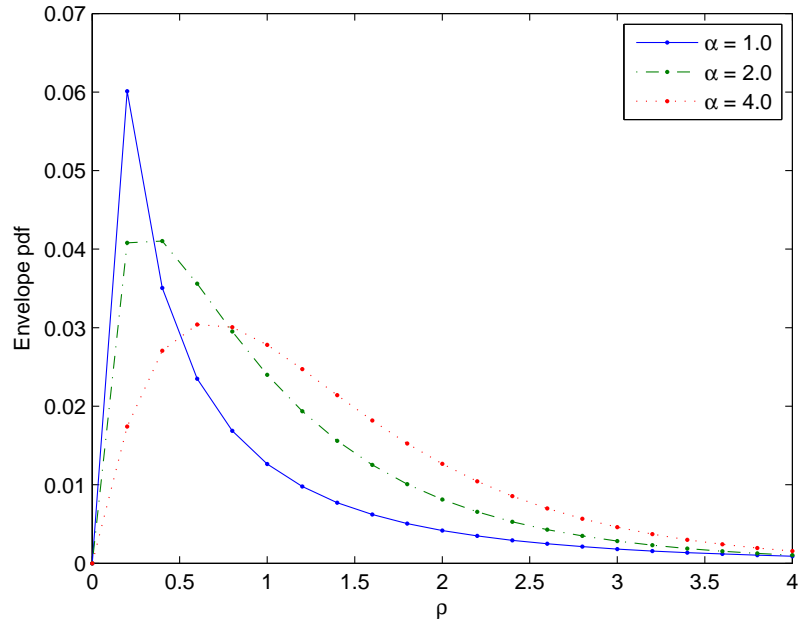,width=12cm, height=9.0cm}}
\caption{Envelope pdf of the $\alpha{-}\kappa{-}\mu{/}$gamma distribution for $\Omega=0.8$, $b = 1.2$, $m = 1.1$ and different values of $\alpha$.}
\end{figure}

\section{Closing Remarks}

This work was devoted to the introduction, formulation and derivation of the $\alpha{-}\kappa{-}mu{/}$gamma composite fading distribution. By using this as a basis, the $\alpha{-}/\mu{/}$gamma and the $\alpha{-}\kappa{-}\mu$ \textit{extreme}${/}$gamma distributions where allso formulated and derived.  These models assume gamma distributed shadowing and are particularly flexible including as special cases the Weibull${/}$gamma, Nakagami-$m$ ${/}$gamma, Rayleigh${/}$gamma and Rice${/}$gamma  multipath${/}$shadowing fading models. Novel analytic expressions were derived for the envelope probability density function which can be considered useful in applications related analytical performance evaluation of digital communications over generalized multipath${/}$shadowing environments with non-linearities.  \\

\bibliographystyle{IEEEtran}

\end{document}